\documentstyle[epsfig]{elsart}
\newcommand{\be}{\begin{equation}}  
\newcommand{\ee}{\end{equation}}
\newcommand{\ba}{\begin{eqnarray}}  
\newcommand{\ea}{\end{eqnarray}}

\newcommand{\ewxy}[2]{\setlength{\epsfxsize}{#2}\epsfbox[45 240 420 350]{#1}}

\newcommand{\Pj}{\mbox{I}\!\!\mbox{P}}

\newcommand{\Tr}{\mbox{Tr}\;}

\newcommand{\RI}{\mbox{\scriptsize RI}}

\newcommand{\MSbar}{\overline{MS}}
\newcommand{\bea}{\begin{eqnarray}}
\newcommand{\eea}{\end{eqnarray}}

\newcommand{\putfig}[1]{\vspace{3.0truecm}\ewxy{#1}{120mm}\vspace{2.5truecm}}
\newcommand{\Dslash}{D\kern-7.8pt\Big{/}}
\newcommand{\pslash}{p\kern-7.8pt\Big{/}}
\newcommand{\qslash}{q\kern-7.8pt\Big{/}}
\begin{document}

%{{{ titles, abstract etc
\rightline{Edinburgh 97-15}
\rightline{FTUV/98-1}
\rightline{IFIC/98-1}
\rightline{Rome Preprint 1182/97}
\rightline{SNS/PH/1998-001}
\begin{frontmatter}

\title{Lattice quark masses: a non-perturbative measurement.}

\author{V.~Gim\'enez}
\address{Dep. de Fisica Teorica and IFIC, Univ. de Val\`encia,\\
Dr. Moliner 50, E-46100, Burjassot, Val\`encia, Spain}
\author{L.~Giusti}
\address{Scuola Normale Superiore, P.zza dei Cavalieri 7 - I-56100 Pisa Italy\\ 
         INFN Sezione di Pisa I-56100 Pisa Italy}
\author{F.~Rapuano}
\address{Dipartimento di Fisica, Universit\`a di Roma \lq La Sapienza\rq ~and\\
         INFN, Sezione di Roma, P.le A. Moro 2, I-00185 Roma, Italy.}
\author{M.~Talevi}
\address{Department of Physics and Astronomy, University of Edinburgh\\
         The King's Buildings, Edinburgh EH9 3JZ, UK.}

\begin{abstract}
We discuss the renormalization of different definitions of quark masses in the 
Wilson and the tree-level improved SW-Clover fermionic action.
Using perturbative and non-perturbative renormalization constants,
we study quark masses in the $\MSbar$ scheme from Lattice QCD in the quenched
approximation at $\beta=6.0$, $\beta=6.2$ and $\beta=6.4$ for both actions.
A reliable extrapolation to the continuum limit is however not yet possible.
The most reliable results are:  
$\overline{m}^{\MSbar}(2\, GeV)=5.7 \pm 0.1 \pm 0.8\, MeV$, 
$m_s^{\MSbar}(2\, GeV)=130 \pm 2 \pm 18\, MeV$ and
$m_c^{\MSbar}(2\, GeV)=1662\pm 30\pm 230\, MeV$. 
\end{abstract}

\end{frontmatter}

\vfill
\centerline{PACS: 11.15.H, 12.38.Gc, 13.30.Eg, 14.20.-c and 14.40.-n}

\newpage
\clearpage
\section{Introduction}
Quark masses are among the least known fundamental parameters of the Standard
Model. Due to confinement, they cannot be measured directly and our knowledge
of these quantities relies on techniques like Chiral perturbation theory (ChPT)
\cite{leut1}, QCD Sum Rules (QCDSR) \cite{derafael}-\cite{dominguez} and
Lattice QCD (LQCD) \cite{qmass1}-\cite{massshill}. ChPT gives rather precise
determinations of ratios of quark masses while  QCDSR and LQCD determine their
absolute values.  Moreover LQCD does not require model parameters or ad-hoc
assumptions.  Each technique  suffers from different sources of errors that
should be carefully studied.
In the most recent LQCD simulations the main errors are due to the  quenched
approximation (i), the reach of the continuum limit (ii) and a correct matching
of the lattice quantities to the continuum ones (iii).
In this paper we try to overcome as much as possible
the last problem. We define quark masses through the Vector and Axial Vector 
Ward Identities and discuss the renormalization procedure.
We compute light, strange and charm quark masses with perturbative (PT) and
non-perturbative (NP) renormalization constants (RC) using the Wilson and the
tree-level improved Sheikholeslami-Wohlert (SW) Clover action \cite{swol},
 \cite{heatlie} which, in the following, we will be referring to as Improved
Action. We also try to estimate  the remaining overall systematic uncertainty,
mainly due to (i) and (ii), comparing the chiral behaviour of the pseudoscalar
and vector meson masses with the experimental ones.

The main result of this paper is a new measurement of the quark masses which we
believe to be more reliable than previous ones.

The paper is organized as follows. In section \ref{sec:qm} we discuss the 
theoretical definitions of lattice quark masses and the renormalization 
procedures. In section \ref{sec:matching} we relate lattice quantities to 
the continuum ones and we review the NP renormalization of bilinear operators.
In section \ref{sec:details} we give 
the details of the lattices used to extract masses and matrix elements.
In section \ref{sec:results} we report our results on quark masses in 
the $\overline{MS}$-scheme and finally we give our conclusions. 

\section{Quark Masses}\label{sec:qm}
The usual on-shell mass definition cannot be used for quarks since they do not
appear as physical states. Thus the values of the quark masses depend on 
the definition adopted. In the following we
will give results for quark masses in the $\overline{MS}$ scheme.

The quark mass can be defined by the perturbative expansion of the
quark propagator renormalized at a scale $\mu$. This is equivalent to the 
definition from the renormalized Vector and Axial Ward Identities,
which can be used to give a fully non-perturbative determination of the quark
mass.

\subsection{Quark masses from the Vector Ward Identity}
Quark masses can be defined from the Vector Ward Identity (VWI). We first
consider the Wilson formulation and then describe the Clover Improved case.
The Vector Ward Identity between on-shell hadronic states $\alpha$ and $\beta$,
neglecting terms of $O(a)$, can be written as \cite{bocax}
\begin{equation}
\langle\alpha|\partial^\mu \tilde V_\mu^a|\beta\rangle =
\langle\alpha|\overline\psi\left[\frac{\lambda^a}{2},M\right]\psi|\beta\rangle\
\end{equation}

where $\psi$ is a flavour triplet of bare quark fields $q_i$, $M$ is the quark
bare mass matrix, $\lambda^{a}/2$ are the generators of the $SU(3)$ flavour 
group normalized by $Tr[\lambda^{a}\, \lambda^{b}] = 2\, \delta^{a b}$ and 
the derivative $\partial ^{\mu}$ is the usual asymmetric lattice
derivative. The conserved vector current $\tilde{V}_{\mu}^{a}$ is:

\begin{eqnarray}
\label{eq:conscurr}
\tilde{V}_{\mu}^{a}(x) &=& \frac{1}{2}\, \left[\, \bar{\psi}(x)\, (\gamma_{\mu}
- 1)\, U_{\mu}(x)\, \frac{\lambda^{a}}{2}\, \psi(x+\mu)\right. \nonumber\\
&+&  \left. \bar{\psi}(x+\mu)\, (\gamma_{\mu} + 1)\, U^{\dagger}_{\mu}(x)\, 
\frac{\lambda^{a}}{2}\, \psi(x)\, \right] 
\end{eqnarray}
and its renormalization constant satisfies $Z_{\tilde{V}}=1$. 
$U_{\mu}$ are the link variables on the lattice. By
taking the linear combination $\tilde V_{\mu}=\tilde V_{\mu}^1-i \tilde V_{\mu}^2$ 
we obtain:
\begin{equation}
\label{eq:vwi}
\langle\alpha|\partial^\mu \tilde V_\mu|\beta\rangle =
\frac{1}{2}\Bigl(\frac{1}{k_2}-\frac{1}{k_1}\Bigr)   
            \langle\alpha| S |\beta\rangle\;
\end{equation}

where $k_i$ is the hopping parameter of $i$-th quark,
$S=\overline q_2 q_1$ is the bare scalar density.

On the lattice the ratio of the vector and scalar matrix elements 
could be used to determine quark masses. Unfortunately the scalar matrix
element turns out to be extremely noisy, preventing any reliable
analysis.  We then use eq.(\ref{eq:vwi}) only to fix the relation  between
the lattice bare quark mass in lattice units and the hopping parameter, i.e.\ 
\be
m_i = \frac{1}{2}\Bigl(\frac{1}{k_i}-\frac{1}{k_c}\Bigr)
\label{eq:mass}
\ee
where $k_c$ is the critical value of the hopping parameter.
Eq. (\ref{eq:vwi}) shows that the quark mass, defined in (\ref{eq:mass}),
is renormalized with $Z_S^{-1}=Z_m$, where $Z_S$ is the renormalization
constant of the scalar density. 
For the pseudoscalar and vector meson masses as a function of 
quark masses, ChPT gives
\be
M_{PS}^2  =  C (m_{1} + m_{2} ) \label{eq:cpt_pio}\; ,
\ee
\be
M_V  =  A + B ( m_{1} + m_{2} )\label{eq:cpt_rho}\; .
\ee
From equations (\ref{eq:cpt_pio}) and (\ref{eq:cpt_rho}) one can 
extract the bare lattice quark mass of a given flavour.
This is the so called ``spectroscopy method". 

In the Improved case eq. (\ref{eq:vwi}) will be more complicate as other terms
will appear due to our use of  ``$\Dslash$-rotated operators" \cite{heatlie}:
\be
q_{i} \longrightarrow \left\{1\, -\, \frac{1}{2}\, \left[\, z\, 
\Dslash\, -\, (1-z)\, m_{i}\, \right]\right\}\, q_{i}
\ee
where $\Dslash$ is the lattice covariant derivative, 
$i$ denotes the flavour of the quark field 
and $z$ is an arbitrary real number. For $z=0$, each field $q_i$ is multiplied 
by a numerical factor so that the tree--level improved bilinear operators
are simply \cite{gabrielli}
\be
O_\Gamma^{I}=(1+ m_i)O_\Gamma\; ,\label{eq:opimpr} 
\ee
obtaining, as for eq. (\ref{eq:vwi}),
\ba
\label{eq:Ivwi}
\langle\alpha|\overline\partial^\mu \tilde V_\mu^{I}|\beta\rangle
&=&\\
\frac{1}{4} (m^I_2 - m^I_1) & &  \langle\alpha|(2 S^{I}(x)+S^{I}(x+\hat0)
+S^{I}(x-\hat0)|\beta\rangle\; ,\nonumber 
\ea
with the symmetric derivative $\overline\partial_\mu f(x) = (f(x+\mu)-f(x-\mu))
/2a$ \cite{vittorio}, where $a$ is the lattice spacing.

However, one should note that the combination of scalar densities in the  right
hand side of eq. (\ref{eq:Ivwi}) can be replaced by $S^{I}(x)$ up to higher 
orders in the lattice spacing $a$. From eqs. (\ref{eq:opimpr})-(\ref{eq:Ivwi})
and the fact  that the Clover Improved action has only  $O(a g^2)$ errors, we obtain
that the mass $m^I_i$, improved up to $O(a g^2)$, is
\be
m^I_i = \frac{1}{2}\Bigl(\frac{1}{k_i}-\frac{1}{k_c}\Bigr) \Bigl[1 - \frac
{1}{4}\Bigl(\frac{1}{k_i}-\frac{1}{k_c}\Bigr)\Bigr]\; .
\label{eq:massI}
\ee
Eq. (\ref{eq:massI}) is, of course, the tree level approximation of the relation
obtained in \cite{luesch2}.
This procedure can be straightforwardly applied to the
perturbative case using RC's that do not contain the field ``rotation''
contribution, but not in the non perturbative one as, for historical
reasons, we have computed RC's using ``rotated'' fields with $z=1$.
 
\subsection{Quark masses from the Axial Ward Identity}
In this section we will describe the definition of the mass from the Axial
Ward Identity (AWI). As for the VWI case we first consider the Wilson
formulation and then describe the Clover Improved case. 
Close to the chiral limit and neglecting terms of $O(a)$, the Axial 
Ward Identity is \cite{bocax}:
\begin{equation}
\label{eq:awi}
Z_A \langle\alpha|\partial^\mu A_\mu^a|\beta\rangle = 
\langle\alpha|\bar \psi\Big\{\frac{\lambda^a}{2},M-\overline{M}\Big\}\gamma_{5}
\psi|\beta\rangle\; , 
\end{equation}
where $Z_A$ is the RC of the axial current and $\overline{M}$,  
defined in \cite{bocax}, in the chiral limit is the critical value of the quark
mass.
Following the same steps as for the VWI, and for the same
flavour choice as in eq. (\ref{eq:vwi}), we obtain that one can
extract quark masses from the ratio of matrix elements
\begin{equation}\label{eq:ratAx}
\frac{1}{2}(\rho_1 + \rho_2) = \frac{\langle 0 | \partial_0 A_0|PS (\vec{p}=0)\rangle}
               {\langle 0 | P|PS (\vec{p}=0)\rangle}\; , 
\end{equation}
where $A_\mu=\overline q_2\gamma_{\mu}\gamma_{5} q_1$, $P=\overline
q_2\gamma_{5} q_1$ and $PS$ is a pseudoscalar meson. This ratio
is renormalized with $Z_A/Z_P$, where $Z_P$ is the RC of the
pseudoscalar density.

For the Clover improved case, the only difference is that we need to
consider the ratio of improved operators and to take the symmetric
derivative, as in eq. (\ref{eq:Ivwi}).

\section{Renormalization of quark masses}\label{sec:matching}

In the continuum, ''physical'' quantities like for example quark masses and
decay constants are defined specifying the renormalization
prescription used to eliminate the divergences in the matrix elements
of the operators determining them. The standard practice consists
in choosing a Dimensional Regularization scheme, for instance,
$NDR-\overline{MS}$.
On the lattice, however, the matrix elements of the corresponding operators are 
regularized by introducing a hard cut--off, $a^{-1}$, in the momentum 
integrals.
Since the continuum and lattice renormalization schemes, in general, differ,
so do the corresponding amplitudes. Therefore, in order to
convert the bare lattice values of amplitudes to their continuum
counterparts, we have to know the matching coefficient, the so--called 
lattice renormalization constant. 

One method, which is described in section 3.1, consists in computing two
amplitudes to a given order in perturbation theory in the continuum and on the
lattice. Imposing that they must be equal after renormalization at a given
scale, one finds the lattice RC's to some order in perturbation theory.  

However, to avoid the use of perturbation theory on the lattice,  a general
method for the non--perturbative calculation of lattice RC's in the RI
(regularization independent) scheme has  been devised. We shall briefly discuss
this in section 3.2. Once we have obtained  the lattice RC's in the RI scheme
non--perturbatively, the conversion to other continuum schemes like
$NDR-{\overline{MS}}$ and the running of the continuum quantity to other scales
can be calculated in perturbation theory at a given order.  

\subsection{The perturbative approach}
In the perturbative approach one uses lattice and continuum 
perturbation theory at the next-to-leading order (NLO).

For the VWI definition, we have 
\ba
m^{\overline{MS}}(\mu) & = & U_m^{\overline{MS}}(\mu,\pi/a)
\frac{1}{Z_S^{\overline{MS}}} m a^{-1} \nonumber\\
& = & U_m^{\overline{MS}}(\mu,\pi/a)  
\left[1+ \frac{\alpha_s(\pi/a)}{4\pi}K_{VWI}\right] m a^{-1}
\ea
where $m$ is the bare quark mass of a generic flavour,
\be\label{eq:kvwi}
K_{VWI} = - C_F ( \Delta_S + \Delta_\Sigma ) - \gamma^{(0)} \log (\pi)\; , \\
\ee
$C_F = (N^2-1)/2N$ and $N$ being the number of colours.
$\Delta_S$ and $\Delta_\Sigma$ have been computed in 
ref. \cite{gabrielli,zhang} and following them can be
extracted from table \ref{tab:matching_pert} for both actions.
\be
U_m^{\overline{MS}}(\mu,\pi / a ) =  \Bigl(\frac{\alpha_s(\mu)}
                       {\alpha_s(\pi / a)}\Bigr)^{\gamma^{(0)}/2\beta_0}\left[
      1+ \frac{\alpha_s(\mu)-\alpha_s(\pi / a)}{4\pi}
\Bigl(\frac{\gamma^{(1)}}{2\beta_0}-\frac{\gamma^{(0)}\beta_1}
{2\beta_0^2}\Bigr)\right]
\ee
where the coupling constant is defined in the $\MSbar$ scheme,
\ba
\beta_0  & = &\frac{11N-2n_f}{3}\nonumber\\
\beta_1  & = &\frac{34}{3}N^2 -\frac{10}{3}N n_f -\frac{N^2-1}{N}n_f\nonumber\\
\gamma^{(0)} & = &6\frac{N^2-1}{2 N} \\
\gamma^{(1)} & = &\frac{97N}{3}\frac{N^2 -1}{2N} + 3\Bigl(\frac{N^2-1}
{2N}\Bigr)^2 -
              \frac{10n_f}{3}\frac{N^2-1}{2N} \nonumber\\
\ea
and $n_f$ is the flavour number.
In the improved case, as discussed before, $\Delta_S$ in eq.~(\ref{eq:kvwi}) 
should not include the contribution due to the rotation of the fields. 

Analogously, for the AWI method, we have
\ba
m^{\overline{MS}}(\mu) & = & U_m^{\overline{MS}}(\mu,\pi/a)
\frac{Z_A^{\overline{MS}}}{Z_P^{\overline{MS}}} \rho a^{-1} \nonumber\\
& = & U_m^{\overline{MS}}(\mu,\pi/a)  
\left[1+ \frac{\alpha_s(\pi/a)}{4\pi}K_{AWI}\right]\rho a^{-1}\; ,
\ea
where $\frac{\rho}{2}$ is the ratio defined in (\ref{eq:ratAx}) for a generic
flavour and
\be\label{eq:avwi}
K_{AWI} = C_F ( \Delta_A - \Delta_P ) - \gamma^{(0)} \log (\pi)\; . \\
\ee
Also $\Delta_A$ and $\Delta_P$ 
can be computed from table \ref{tab:matching_pert} following 
\cite{gabrielli,zhang}.

\subsection{The non-perturbative approach}\label{sec:np}
At scales $a^{-1}\simeq 2-4$ GeV, where $a$ is the lattice spacing, of our 
simulations, we expect small NP effects on the renormalization constants of
bilinear operators. However
``tadpole'' diagrams \cite{makenzie}, which are present in lattice perturbation
theory, can give rise to large corrections and then to large
uncertainties in the matching procedure at values of $\beta=6/g_L^2=6.0-6.4$.
These problems are avoided using NP renormalization techniques
\cite{NPM,luescher}. Although the nature of the systematics changes in adopting
a NP approach, mainly because of $O(a)$ effects as we shall discuss in 
\cite{Z_bl_tli}, we believe that the NP determination of RC's is more reliable.

We shortly review the NP method for the RC's which enter 
the determination of the quark masses. For the full discussion of 
the method, the results obtained and their systematics we refer to a 
forthcoming paper \cite{Z_bl_tli}.

Let us consider a quark bilinear 
$O_{\Gamma}=\overline q \Gamma q$, where $\Gamma$ is a Dirac matrix
and $q$ now represents the generic lattice flavour-degenerate 
quark field.
In this work we will consider the scalar and pseudoscalar
densities and the Axial Vector current.

The renormalization conditions are imposed on the 
amputated Green functions computed between off-shell quark states of 
momentum $p$ in the Landau gauge
\begin{equation}
\Lambda_O(pa)=S_q(pa)^{-1}G_O(pa)S_q(pa)^{-1}
\end{equation}  
where $G_O(pa)$ and $S_q(pa)$ are the non-amputated Green functions and
the quark propagator, calculated non-perturbatively via Monte Carlo simulations
\cite{NPM}. Possible effects from Gribov copies or spurious solutions 
\cite{tesi_mia} have not been considered.  
The RC $Z^{\RI}_O(\mu a, g_0)$ of $O_\Gamma$
is determined by the condition
\begin{equation}
Z_O^{\RI}(\mu a)Z_q^{-1}(\mu a) \Tr\Pj_O\Lambda_O(pa)|_{p^2=\mu^2}=1,
\label{eq:RI}
\end{equation}
where $\Pj_O$ is a suitable projector on the tree-level amputated
Green function (normalized to 1) \cite{NPM} and $Z_q$ is the wave function
RC which can be defined in different ways \cite{NPM}.
From the Ward Identities 
\begin{equation}
Z_q(\mu a)=\left.-i\frac{1}{12}Tr\Bigl(
\frac{\partial S_q(pa)^{-1}}{\partial\pslash}\Bigr)\right|_{p^2=\mu^2}\; .
\end{equation}
To avoid derivatives with respect to a discrete variable, we have used  
\begin{equation}
Z'_q(\mu a)=\left.-i\frac{1}{12}\frac{Tr\sum_{\mu=1,4}\gamma_\mu \sin(p_\mu
a)S_q(pa)^{-1}}{4\sum_{\mu=1,4} \sin^2(p_\mu a)}\right|_{p^2=\mu^2}\; ,
\label{eq:Z_q_WI}
\end{equation}
which, in the Landau Gauge, differs from $Z_q$ by a finite term of order
$\alpha_s^2$ which is irrelevant at the NLO. 

For this procedure to be reliable $\mu$ must satisfy the condition 
$\mu \ll 1/a$ to avoid discretization errors but also $\mu\gg \Lambda_{QCD}$ 
to avoid non-perturbative effects or higher order corrections in 
the continuum perturbative expansion.

The quark mass in the $\overline{MS}$ scheme is then defined as
\be\label{eq:matching}
m^{\overline{MS}}(\mu) =  U_m^{\overline{MS}}(\mu,\mu')
\left[1+ \frac{\alpha_s(\mu')}{4\pi}C^{LAN}_m\right]
m^{RI}(\mu')\;,
\ee
where \cite{NPM}
\be
C_m^{LAN} =  - 4 \frac{N^2-1}{2N}
\ee
and for $m^{RI}(\mu')$ we get
\ba
m^{RI}(\mu') & = & \frac{1}{Z_S^{RI}(\mu' a)} m a^{-1}\nonumber\\
m^{RI}(\mu') & = & \frac{Z_A^{RI}}{Z_P^{RI}(\mu' a)}\frac{\rho}{2} a^{-1}\; 
\ea
for the Vector and Axial Vector WI respectively. As both the $RI$ and the
$\overline{MS}$ respect chirality we have $Z_A^{RI}=Z_A$.

This discussion applies immediately to the unquenched case \cite{tchil}.
  
\setlength{\tabcolsep}{.16pc}
\begin{table}
\begin{center} 
\begin{tabular}{||c||c||c|c||c|c||c|c||}
\hline\hline     %1 2 3 4 5 6
Regularization  & $\Delta_\Sigma$ & $\Delta_S^{Loc}$ & $\Delta_S^{\otimes} $ &
$\Delta_P^{Loc}$ & $\Delta_P^{\otimes}$ & $\Delta_A^{Loc}$ & 
$\Delta_A^{\otimes}$\\
\hline
Wilson  & -12.85 & -0.10  &  -    & -9.78 &   -   & -3.0 &   -   \\
Clover  & -9.2   & -10.07 & 12.0  & -13.1 & -4.3  & -4.6 & 11.7  \\
\hline
\hline
\end{tabular}
\end{center}
\vspace{.3truecm}
\caption{Coefficients entering the perturbative matching of the quark masses.
\label{tab:matching_pert}}
\end{table}

\section{Details of the analysis}\label{sec:details}
In this section we describe the extraction of meson masses, matrix
elements and lattice quark masses from the two point correlation functions and
the interpolation/extrapolation of the results in the heavy and light quark 
masses to 
the physical points. We also try to estimate the overall systematic errors 
on the lattice quark masses.

\subsection{Lattice details}
In this work we have used various lattices that have been generated by the APE
group in the last years. Tables~\ref{tab:clo} and \ref{tab:wil} show the
parameters of the lattices that we have analyzed. A more detailed
discussion of lattice calibration and spectroscopy can be found in \cite{mio1}
for the light quark systems and in \cite{mio2} for the heavy-light ones. Here we
will just summarize the most important points.

Meson masses and axial-pseudoscalar matrix elements have been extracted from 
two-point correlation functions in the standard way from the following 
propagators   
\begin{eqnarray}\label{eq:scalare}
G_{55}(t) & = & \sum_{x}\langle P(x,t)P^\dagger(0,0) \rangle\; ,\nonumber\\
G_{05}(t) & = & \sum_{x}\langle A_0(x,t)P^\dagger(0,0) \rangle 
\end{eqnarray}
and 
\be\label{eq:vec}
G_{ii}(t) =  \sum_{i=1,3}\sum_{x}\langle V_i(x,t)V_i^\dagger(0,0)\rangle\; ,
\ee 
where
\begin{eqnarray}
P(x,t)    & = & \overline{q}(x,t)\gamma_5q(x,t)\; ,\nonumber\\
A_\mu(x,t)  & = & \overline{q}(x,t)\gamma_\mu\gamma_5q(x,t)\; ,\nonumber\\ 
V_i(x,t)    & = & \overline{q}(x,t)\gamma_i q(x,t)\; .\nonumber
\end{eqnarray}
where $q$ represents the generic lattice flavour-degenerate quark field
and $i$ is a spatial index.
We fit the zero-momentum correlation functions in eqs.~(\ref{eq:scalare}) 
and (\ref{eq:vec}) 
to a single particle propagator
\begin{eqnarray}
G_{55}(t) & = & \frac{Z^{55}}{M_{PS}}\exp\Bigl(-\frac{1}{2}M_{PS}T\Bigr)
\cosh\Bigl[M_{PS}\Bigl(\frac{T}{2}-t\Bigr)\Bigr]\; ,\nonumber\\
& & \nonumber\\
G_{ii}(t) & = & \frac{Z^{ii}}{M_V} \exp\Bigl(-\frac{1}{2}M_{V}T\Bigr)
\cosh\Bigl[M_{V}(\frac{T}{2}-t\Bigr)\Bigr]\; ,\nonumber\\
& & \label{eq:funzfit}
\end{eqnarray}
in the time intervals reported in tables~\ref{tab:clo}
and \ref{tab:wil}.
In (\ref{eq:funzfit}), $T$ represents the lattice time extension, the
subscripts $PS$ and $V$ stand for pseudoscalar and vector mesons.  To improve
stability, the meson (axial-pseudoscalar) correlation functions have been
symmetrized (anti-symmetrized) around $t=T/2$.  The time intervals for the fits
are chosen with the following criteria: we fix the lower limit of the interval
as the one at which there is a stabilization of the effective  mass,  and, as
the upper limit, the furthest possible point before the error overwhelms the
signal. 
%--------------------------------------------------------------
\setlength{\tabcolsep}{.16pc}
\begin{table}
\begin{center} 
\begin{tabular}{||c|cccccc||}
\hline\hline       %123456
&C60a&C60b&C60c&C60d&C62a&C64\\
\hline
$\beta$&$6.0$&$6.0$&$6.0$ &$6.0$&$6.2$&$6.4$\\
\# Confs&490&600&200&200&250&400\\
Volume&$18^3\times 64$&$24^3\times 40$&$18^3\times 32$& 
$16^3\times 32$&$24^3\times 64$&$24^3\times 64$\\
\hline
$k_l$&0.1425&0.1425&0.1425&0.1425&0.14144&0.1400\\
     &0.1432&0.1432&0.1432&0.1432&0.14184&0.1403\\
     &0.1440&0.1440&0.1440&0.1440&0.14224&0.1406\\
     &  -   &   -  &   -  &   -  &0.14264&0.1409\\
\hline
$k_h$&  -   &  -   &  -   &  -   &0.1210& - \\    
     &  -   &  -   &  -   &  -   &0.1250& - \\    
     &  -   &  -   &  -   &  -   &0.1290& - \\    
     &  -   &  -   &  -   &  -   &0.1330& - \\    
\hline
& \multicolumn{5}{c}{Light-light mesons}&\\
$t_1 - t_2$ & 15-28 & 15-19 & 11-15 & 11-15 & 18-28 & 24-30\\
\hline
& \multicolumn{5}{c}{Heavy-light mesons}& \\
$t_1 - t_2$ &   -   &  -    &   -   &   -   & 20-28 & - \\
\hline
\hline
$a^{-1}(K^*)$& 2.12(6) & 2.16(4) & 2.07(6) & 2.23(9) & 2.7(1) & 4.0(2)\\  
\hline
\hline
\end{tabular}
\end{center}
\vspace{.3truecm}
\caption{Summary of the parameters of the runs with the SW-Clover fermion 
action analyzed in this work.\label{tab:clo}}
\end{table}
%-------------------------------------------------- 
%------------------------------------------------
\setlength{\tabcolsep}{.16pc}
\begin{table}
\begin{center} 
\begin{tabular}{||c|cccc||}
\hline\hline       %1234
&W60&W62a&W62b&W64\\
\hline
$\beta$&$6.0$&$6.2$&$6.2$&$6.4$\\
\# Confs&320&250&110&400\\
Volume&$18^3\times 64$& $24^3\times 64$ 
&$24^3\times 64$&$24^3\times 64$\\
\hline
$k_l$&0.1530&0.1510&0.1510&0.1488\\
     &0.1540&0.1515& -    &0.1492\\
     &0.1550&0.1520&0.1520&0.1496\\
     &  -   &0.1526&0.1526&0.1500\\
\hline
$k_h$&0.1255&0.1300&0.1300&  -   \\    
     &0.1320&0.1350&0.1350&  -   \\    
     &0.1385&0.1400&0.1400&  -   \\    
     &0.1420&0.1450&0.1450&  -    \\    
     &0.1455&   -  &0.1500&  -    \\     
\hline
& \multicolumn{3}{c}{Light-light mesons}& \\
$t_1 - t_2$ & 15-28 & 18-28 & 18-28 & 24-30\\
\hline
& \multicolumn{3}{c}{Heavy-light mesons}& \\
$t_1 - t_2$ & 15-28 & 20-28 & 20-28 &  - \\
\hline
$a^{-1}(K^*)$& 2.26(5)& 3.00(9) & 3.0(1) & 4.1(2) \\  
\hline
\hline
\end{tabular}
\end{center}
\vspace{0.3truecm}
\caption{Summary of the parameters of the runs with the Wilson action 
analyzed in this work.\label{tab:wil}}
\end{table}
%----------------------------------------------- 
The errors have been estimated by a jacknife procedure, blocking the data in
groups of 10 configurations and we have checked that there are no relevant 
changes in the error estimate by blocking groups of
configurations of different size.\\ 
We extract $\rho$ from the ratio 
\be
\rho = 
\frac{\langle \partial_0 A_0(t) P(0) \rangle}
     {\langle P(t) P(0)\rangle}
\ee
at zero spatial momentum. At large time separations, we fit the ratio
$G_{05}(t)/G_{55}(t)$ and we get
\be\label{eq:lwi}
\rho = \sinh(M_{PS})
\frac{\langle A_0(t) P(0) \rangle}
     {\langle P(t) P(0)\rangle}
\ee
in a $t$ region where the signal stabilizes. The hyperbolic sine comes from the
discrete symmetric derivative which is necessary to consistently take into
account $O(a)$ terms.

Tables \ref{tab:clo} and \ref{tab:wil} also contain the values of the 
lattice spacing $a$ extracted from the $K^*$ mass with the "lattice physical 
plane" (lp-plane) method \cite{mio1}.

We have also performed dedicated runs at $\beta=6.0, 6.2$ and $6.4$ for both the Wilson
and the SW-Clover actions to measure the RC's according to the
discussion of section \ref{sec:np}.
The parameters of these runs and the results for $Z_S^{RI}$, $Z_P^{RI}$ and 
$Z_A^{RI}$ are reported in table \ref{tab:ZPZA}.

\setlength{\tabcolsep}{.16pc}
\begin{table}[htb]
\begin{center}
\begin{tabular}{||l|c|c|c|c|c|c||}
\hline\hline
$\beta$ & 6.0  & 6.0    & 6.2 & 6.2    & 6.4 & 6.4  \\
\hline
Action   & SW   & Wilson & SW  & Wilson & SW  & Wilson \\
\# Confs & 100  & 100    & 180 & 100    & 60  & 60     \\
Volume   &$16^3\times 32$ & $16^3\times 32$ & $16^3\times 32$ & $16^3\times 32$
         &$24^3\times 32$ & $24^3\times 32$ \\
\hline
$k$ &0.1425& 0.1530 & 0.14144 & 0.1510 & 0.1400 & 0.1488 \\
         &0.1432& 0.1540 & 0.14184 & 0.1515 & 0.1403 & 0.1492 \\
         &0.1440& 0.1550 & 0.14224 & 0.1520 & 0.1406 & 0.1496 \\
         &      &        & 0.14264 & 0.1526 & 0.1409 & 0.1500 \\
\hline
$k_c$ &0.14551& 0.15683 & 0.14319 & 0.15337 & 0.14143 & 0.15058\\
\hline\hline
$Z_S^{RI}(m_qa=0)$& 0.834(18) & 0.682(9) & 0.851(11)& 0.722(8) & 0.852(13) & 0.742(8)\\
$Z_P^{RI}(m_qa=0)$& 0.409(8)  & 0.447(5) & 0.466(4) & 0.499(5) &  0.555(6) & 0.572(4)\\
$Z_A^{RI}(m_qa=0)$& 1.047(18) & 0.808(7) & 1.023(4) & 0.812(6) & 1.012(9) & 0.825(6) \\
\hline
\hline
\end{tabular}
\end{center}
\vspace{0.3truecm}
\caption{Parameters of the runs used in the non perturbative calculation of 
the renormalization constants and values of $Z_S^{RI}$, $Z_P^{RI}$
and $Z_A^{RI}$ at a scale $\mu a \simeq 1$.
The errors reported are statistical only.\label{tab:ZPZA}}
\end{table}

\subsection{Extraction of raw results for light and strange quarks}
Once the hadronic correlation functions have been fitted, and the lattice
masses and matrix elements extracted, we have to perform a number of
interpolations/extrapolations to extract physical quantities.\\ 
We extract light and strange quark masses from the meson 
spectroscopy and from the Axial Ward Identity with three different methods:
\begin{enumerate}
\item\label{item:1} We define the physical plane $[M_V, \; M_{PS}^2]$
(lp-plane \cite{mio1}) and, assuming that only linear terms are
important, we determine the lattice values $M_K$ and $M_{K^*}$ 
imposing that 
$M_V/M_{PS}$ coincides with the experimental value
$C_{sl}=M^{exp}_{K^*}/M^{exp}_K$ \cite{pdg}.
Comparing $M_{K^*} a^{-1}$ with its experimental value we extract the lattice
spacing reported in tables \ref{tab:clo} and \ref{tab:wil}. 
Using relations (\ref{eq:cpt_pio}) and (\ref{eq:mass}) and imposing 
the prediction of the CPTh \cite{leut1}
\be\label{eq:R}
R = \frac{m_s}{\overline{m}} = 24.4 \pm 1.5\; ,
\ee
where $\overline{m}=(m_u+m_d)/2$, we obtain 
\ba
m_s & = & \frac{M_K^2}{C (1+1/R)}\label{eq:strange1}\\
\overline{m} & = & \frac{m_s }{R} \; .\label{eq:qmass1}
\ea
The lattice strange quark mass obtained with (\ref{eq:strange1}) 
for all lattices are reported in table \ref{tab:quarkm_sp}. 

In extracting the values of $a^{-1}$, $M_K$ and $C$ from our data,
we assume that the relation between quark masses and squared 
pseudoscalar masses is linear up to the strange quark region (see eq.(2)),
so we can compute these quantities using our data for quark masses in
the ``strange mass region'' by interpolating only. In other words, no
extrapolation to the light quark region is performed. 
Possible deviations from linearity induced by terms of the form
$D\, (m_{s} + \overline{m})^{2}\, +\, D'\, (m_{s} - \overline{m})^{2} $
are known to be small and
will give rise to uncertainties which can be neglected. 
Moreover, the value of $k_{c}$ is not
required to obtain the quark masses in the Wilson case. 
For the improved action, however, a small dependence of $C$
on $k_{c}$ exists because of the quadratic term in eq.(7). This gives rise
anyway to quite small fluctuations.

For the Axial Ward Identity, in the same fashion, we define the plane 
$[\rho,\; M_{PS}^2]$, where $\rho$ is defined in the equation (\ref{eq:lwi}),
and we obtain
\ba
\rho_s & = & \frac{\rho|_{M_{PS}=M_K}}{(1+1/R)}\label{eq:rhos}\\
\overline{\rho} & = & \frac{\rho_s}{R}\; . 
\ea
In this case, for both actions, there is no chiral extrapolation needed.
The
values of $\rho_s$ for all lattices are reported in table \ref{tab:quarkm_wi}.
\item\label{item:2} On the physical plane $[M_V\; M_{PS}^2]$, we determine 
$M_K$ and $M_{K^*}$ as in (\ref{item:1}) and $M_\pi$ and  $M_{\rho}$ imposing
that $M_V/M_{PS}$ coincides with the experimental value 
$C_{ll}=M^{exp}_{\rho}/M^{exp}_\pi$. Using (\ref{eq:cpt_pio}) and
(\ref{eq:mass}) we obtain 
\ba
\overline{m} & = & \frac{M_\pi^2}{2 C}\label{eq:light1}\\
m_s   & = & \frac{1}{C} \left(M_K^2 - \frac{M_{\pi}^2}{2}\right)\; .
\ea
The lattice light quark masses obtained with (\ref{eq:light1}) for all
lattices are reported in table \ref{tab:quarkm_sp}.
For the Axial Ward Identity we define the plane 
$[\rho,\; M_{PS}^2]$ and we obtain
\ba
\overline{\rho} & = & \left.\frac{1}{2}\rho\right|_{M_{PS}=M_\pi}\nonumber\\
\rho_s     & = & \left.\rho\right|_{M_{PS}=M_K} - 
   \left. \frac{1}{2}\rho\right|_{M_{PS}=M_{\pi}}\; .
\ea
The values of $\overline{\rho}$ for all lattices are reported in
table \ref{tab:quarkm_wi}.
\item\label{item:3} Using eqs.~(\ref{eq:mass}), (\ref{eq:cpt_rho})  and the 
lattice spacing determined as in 
(\ref{item:1}) we compare $(M_\phi a^{-1})$ to its experimental
value and we obtain
\be\label{eq:phis}
m_s = \frac{M_\Phi^{exp} a - A}{2 B} .
\ee
From (\ref{eq:mass}) and (\ref{eq:phis}) we determine $k_s$ and then 
we use 
\be
\rho = G \Bigl(\frac{1}{k} - \frac{1}{k_c}\Bigr)
\ee 
to determine $\rho_s$. 
\end{enumerate} 
The three methods (\ref{item:1}), (\ref{item:2}) and (\ref{item:3}) 
described above should give 
consistent values for $\overline{m}$, $m_s$, $\overline{\rho}$ 
and $\rho_s$, apart from discretization errors and quenching effects.
It is well known \cite{mio1,guptastronge} that the strange
quark mass obtained 
from $M_{\phi}$ (\ref{item:3}) is systematically higher than the value
obtained from $M_K$ (\ref{item:1}).

The ratio of the strange quark masses obtained 
with different methods is connected to experimental quantities. 
From (\ref{eq:cpt_pio}) and (\ref{eq:cpt_rho}) we get
\be
M_V  =  A + D M_{PS}^2
\ee
where $D\cdot C = B$. $A$ and $D$ can be measured fitting linearly 
the experimental data on the meson masses \cite{pdg}. To compare 
lattice data with experimental measurements we prefer to use
two equivalent dimensionless quantities 
\ba
J & \equiv & M_{K^*}\frac{d M_V}{d M_{PS}^2} = M_{K^*} D\nonumber\\
L & \equiv & \frac{A}{M_{K^*}}\; . 
\ea                       
In table \ref{tab:JLR} we report the experimental values 
and the lattice predictions we have obtained for $J$ and $L$.
One can see that $J$ has quite large fluctuations and is also quite far from the
experimental value. It has been argued \cite{michael} that this may be mainly due
to the quenching approximation.\\

\setlength{\tabcolsep}{.18pc}
\begin{table}[htb]
\begin{center}
\begin{tabular}{||c|ccc||}
\hline\hline
 Run    &$J$    & $L$   & $R_{m_s}$ \\
\hline
Exp.   & 0.499 & 0.847 & 1     \\
\hline
C60a    & 0.367(13)& 0.887(4) & 0.852(17) \\
C60b    & 0.377(8) & 0.884(2) & 0.866(10) \\
C60c    & 0.359(16)& 0.890(5) & 0.842(22) \\
C60d    & 0.393(11)& 0.879(3) & 0.884(14) \\
W60     & 0.339(10)& 0.896(3) & 0.815(14) \\
C62a    & 0.343(50)& 0.895(15)& 0.821(68) \\
W62a    & 0.360(21)& 0.889(7) & 0.843(28) \\
W62b    & 0.380(28)& 0.883(9) & 0.869(35) \\
W64     & 0.386(16)& 0.881(5) & 0.876(19) \\
C64     & 0.401(17)& 0.877(5) & 0.894(20) \\
\hline
\hline
\end{tabular}
\end{center}
\vspace{0.3truecm}
\caption{Experimental and lattice values for $J$, $L$ and $R_{m_s}$.
\label{tab:JLR}}
\end{table}    
The ratio of the quark masses obtained with different methods is
only function of $J$ and $L$ and experimental ratios. In particular the ratio
of the strange quark mass obtained with (\ref{item:1}) and (\ref{item:3}) is
\be\label{eq:rms}
R_{m_s} = \frac{(m_s)_K}{(m_s)_\phi} = \frac{1}{C^2_{sl}(1+\frac{1}{R})}
          \frac{2 J}{\frac{M_\Phi^{exp}}{M_{K^*}^{exp}}-L} \; . 
\ee
where $R$ is defined in eq.~(\ref{eq:R}).
In Table \ref{tab:JLR} we report the experimental value for 
$R^{Exp}_{m_s}$ and the values we have obtained for our lattices. 
$R_{m_s}$ is a ratio of different
definitions of lattice quark masses that does not suffer from $O(a)$ effects
induced by renormalization constants and is related to experimental
quantities.
We believe that the difference 
$R^{Exp}_{m_s} - R^{Lat}_{m_s}$ is a good estimate of the overall 
systematic error on the strange quark mass. This error should essentially 
take into account both the quenching approximation and the 
$O(a)$ effects that we cannot a priori estimate. 

\subsection{Extraction of raw results for the charm quark mass}
In order to obtain the charm quark mass we have to extrapolate the meson masses
and the matrix elements both in the heavy and light quark masses. 
It is clear that in this case $O(ma)$ effects will be much more relevant than
for the strange and light quarks. We use two different methods:
\begin{enumerate}
\item \label{pesante:1}We first interpolate/extrapolate linearly 
the heavy-light pseudoscalar meson masses in the light quark 
mass to $m_s$ extracted from (\ref{eq:strange1}). We then determine the
charm quark mass by fixing the $D_s$-meson mass to its physical value and 
using the equation (\ref{eq:mass}). The lattice charm quark masses obtained 
in this way for all lattices are reported in table 
\ref{tab:quarkm_sp}. For the Axial Ward Identity $\rho_c$ is given by
\be
\rho_c = \rho |_{M_{PS}=M_{D_s}} - \rho_s
\ee 
with $\rho_s$ defined in eq.~(\ref{eq:rhos}). The values obtained for all 
lattices are reported in table \ref{tab:quarkm_wi}.
\item As in (\ref{pesante:1}) but we determine the value of the charm quark
mass by fixing the $D^*_s$-meson mass to its physical value. 
\end{enumerate}
As for the light quark masses, the two methods described above should give 
consistent results for $m_c$ and $\rho_c$, 
apart from discretization errors and quenching effects. We shall use the spread
between the two determinations as an estimate of the overall systematic error.

\section{Physical Results for quark masses}\label{sec:results}
On the basis of the discussion of previous sections, we now present the final
results.
From the lattice data of the left columns of tables \ref{tab:quarkm_sp} and
\ref{tab:quarkm_wi}, we obtain the PT and NP $\MSbar$ results reported
in the same tables.  

In the PT case there is a large ambiguity in the choice among the lattice strong
coupling constants which have been proposed to have a better convergence of the
series. In the appendix we give in detail which definitions we have used. The
difference among them is of course $O(\alpha_s^2)$ but on the lattice the
behaviour of the perturbative series can be very bad \cite{makenzie}. The
different values are averaged and the spread taken as an estimate of the error.
\par In the continuum we have used both in the PT and NP case eq. (\ref{eq:alfstr})
with $n_f=0$ and $\Lambda_{QCD}= 251 \pm 21\; MeV$ \cite{capit} as
discussed in the appendix. We have checked that the final results change by
less then $3\%$ if we had used $n_f=4$ and $\Lambda_{QCD} = 340 \pm 120\; MeV$,
and this is only part of the error due to the quenching approximation which we
cannot estimate.

\begin{figure}[t] 
\putfig{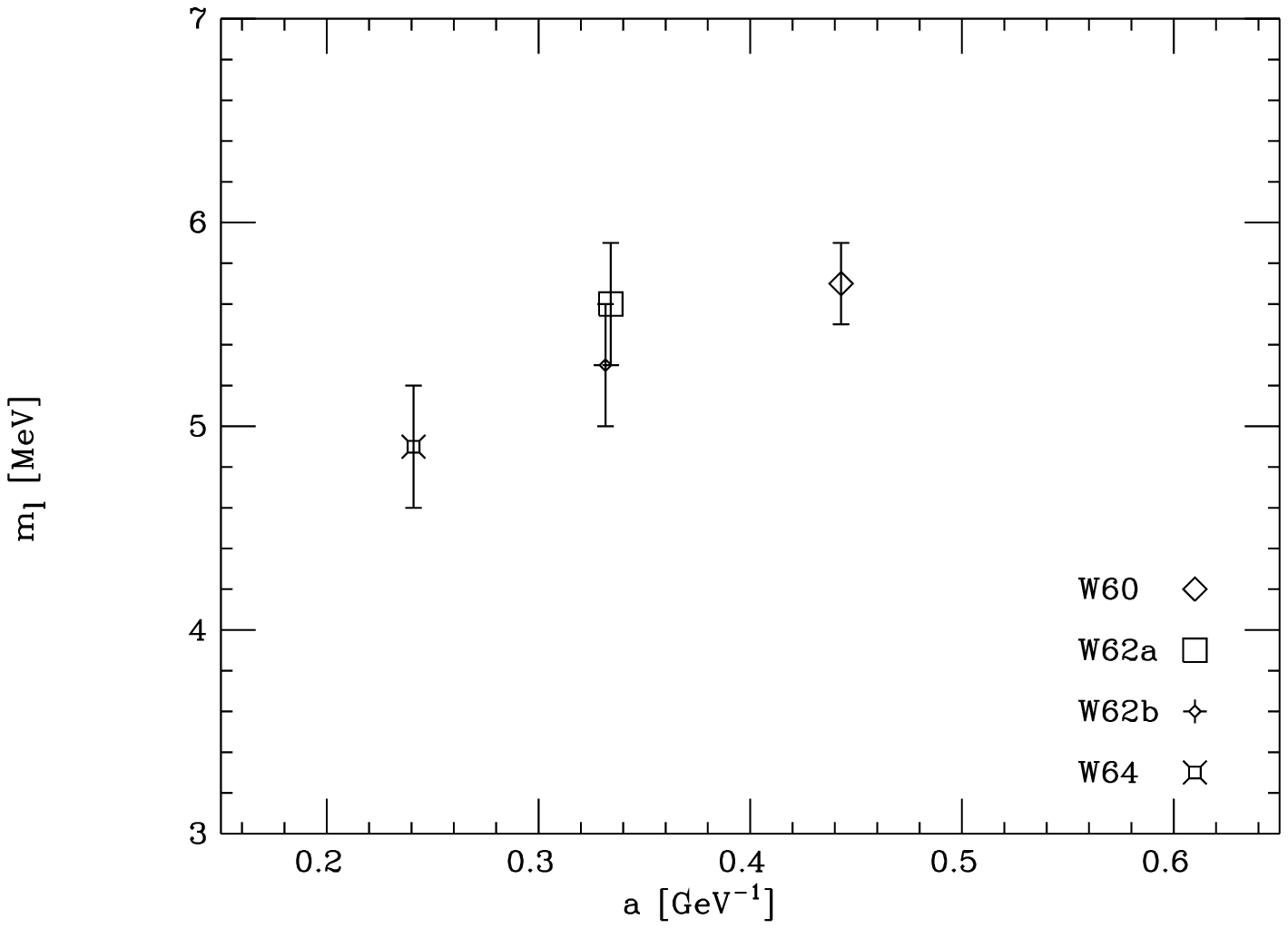}
\vspace{0.3truecm}
\caption[]{Non-perturbatively renormalized quark masses $\overline{m}^{\MSbar}$
for all Wilson lattices from the Axial Ward Identity}
\protect\label{fig:light_wil_awi}
\end{figure}

\begin{figure}[t] 
\putfig{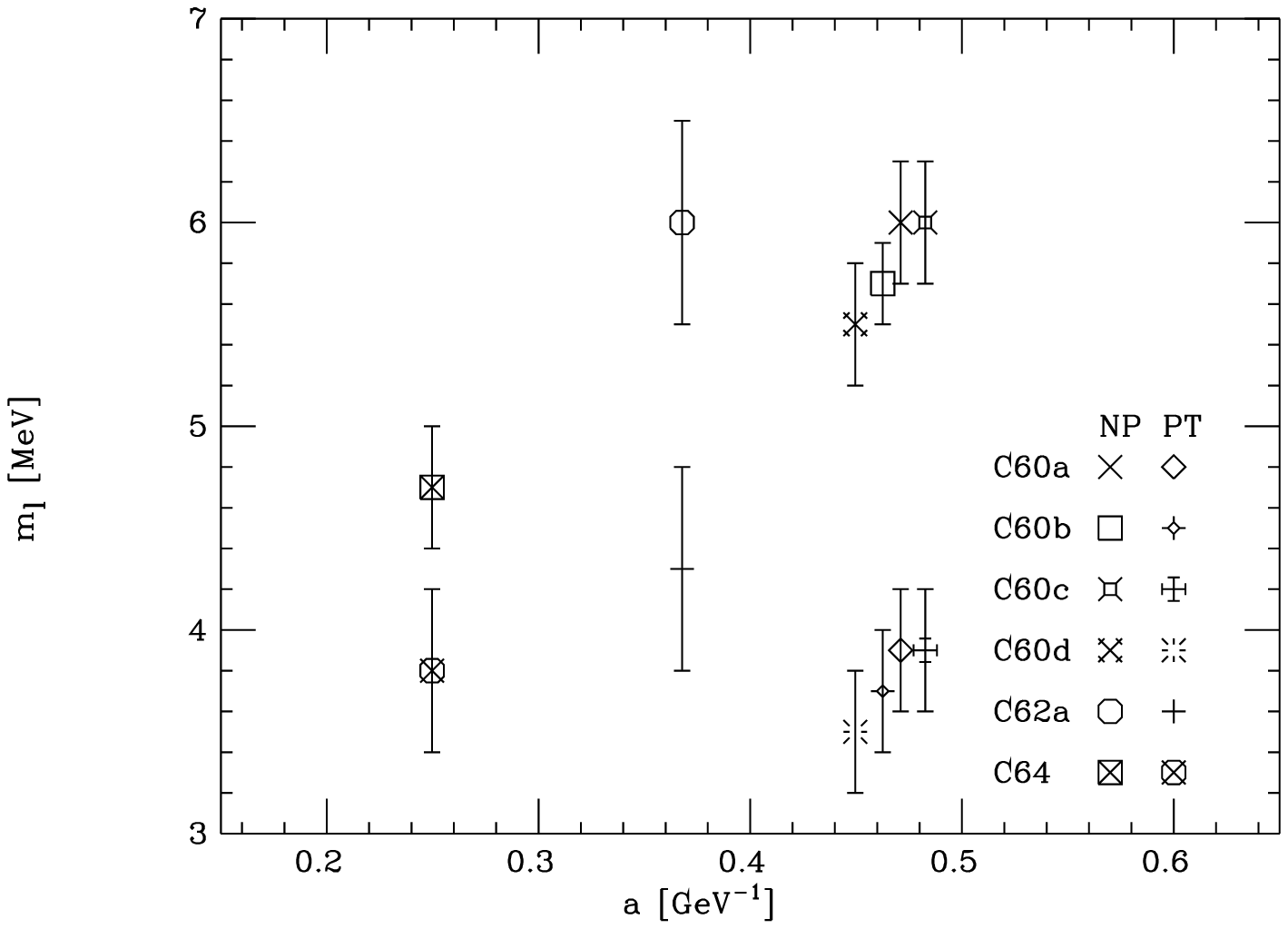}
\vspace{0.3truecm}
\caption[]{Non-perturbatively renormalized quark masses $\overline{m}^{\MSbar}$
for all SW-Clover lattices from the Axial Ward Identity}
\protect\label{fig:light_clo_awi}
\end{figure}

\begin{figure}[t] 
\putfig{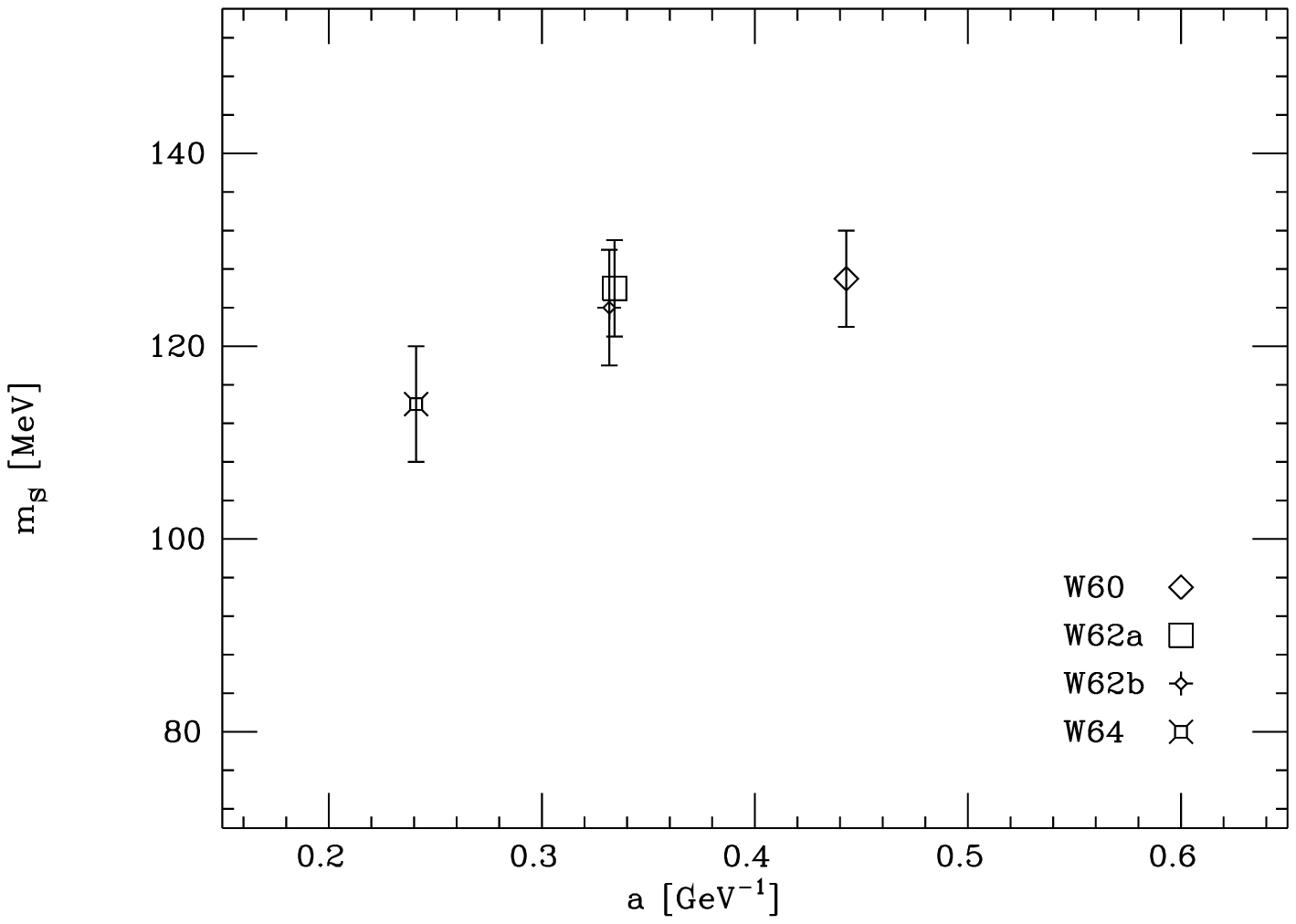}
\vspace{0.3truecm}
\caption[]{Non-perturbatively renormalized quark masses $m_s^{\MSbar}$ for all
Wilson lattices from the Axial Ward Identity}
\protect\label{fig:strange_wil_awi}
\end{figure}

\begin{figure}[htb] 
\putfig{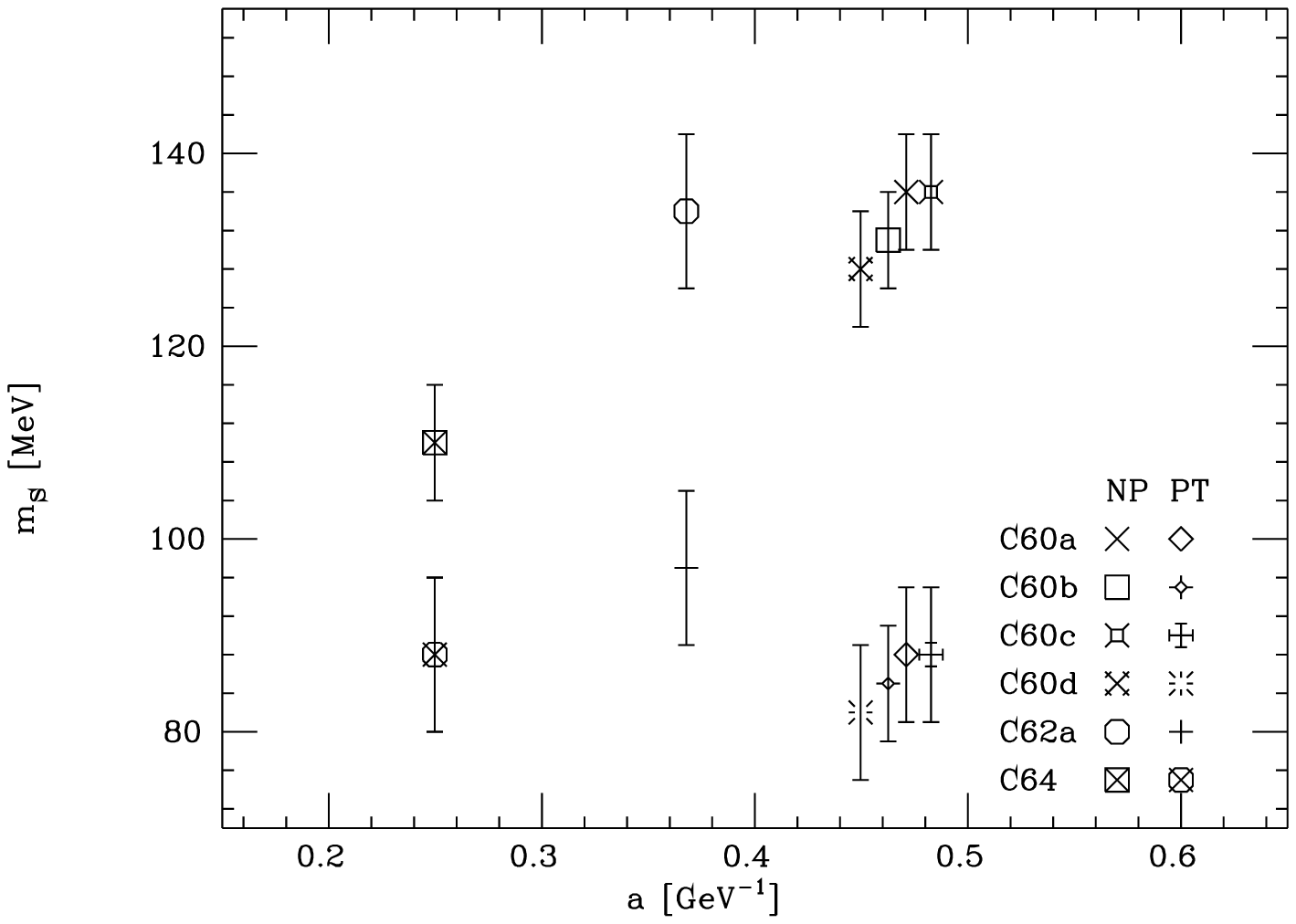}
\vspace{0.3truecm}
\caption[]{Non-perturbatively renormalized quark masses $m_s^{\MSbar}$ for all
SW-Clover lattices from the Axial Ward Identity}
\protect\label{fig:strange_clo_awi}
\end{figure}

The NP results contain only the statistical error on the matrix elements and
renormalization constants. The  AWI results are shown in figs. 
\ref{fig:light_wil_awi}-\ref{fig:strange_clo_awi}.

In the spectroscopy case we find a reasonable agreement between the NP and PT 
results that turn out to be compatible with previous determinations 
\cite{qmass1,mio1,guptastronge}. This is not the case for the AWI where 
the PT results are lower than the NP ones by more than two standard 
deviations. In our opinion this confirms that the perturbation theory
fails in the determination of the pseudoscalar RC. On the other hand the NP 
results for different actions and methods are in very good agreement among them.

As far as the $a$ dependence is concerned, for $\beta =6.0$ and $6.2$ the data
are definitely stable, within the errors, for all methods. At  $\beta = 6.4$
the data apparently show a slight decrease.   Due to the small physical volume
at this $\beta$ value and the fact that this effect is present both in the
Wilson and the Clover case we believe that  one cannot disentangle volume and
$O(a)$ effects in this lattice \cite{mio1} and we  will not take into account
in any further result. With our present data an extrapolation in $a$ is then
out of reach.

The charm quark results appear to be quite more noisy than the light and strange
ones. This supports the fact that at large masses order $am$ contaminations
are important.

To obtain final results, we average the non perturbative AWI results
at $\beta =6.0$ and $6.2$ as independent ones.
For the strange and charm quark, we also take into account the overall
systematic error which we evaluate from the spread in the quark masses
extracted from different mesons as described in sec. \ref{sec:details}. 
This error is also propagated to the light quark using eq. 
(\ref{eq:qmass1}) and is the second one in eq. (\ref{res:np_awi}).

From NP renormalized AWI we obtain our final results
\ba\label{res:np_awi}
\overline{m}^{\MSbar}(2\, GeV) & = & 5.7 \pm 0.1 \pm 0.8\; Mev\nonumber\\  
         m_s^{\MSbar}(2\, GeV) & = & 130 \pm 2  \pm 18\; Mev \\
         m_c^{\MSbar}(2\, GeV) & = & 1662 \pm 30 \pm 230\; Mev \nonumber\; .
\ea
which are reported in the abstract. 
\setlength{\tabcolsep}{.18pc}
\begin{table}[htb]
\begin{center}
\begin{tabular}{||c|cccc|cccc|cccc||}
\hline\hline
 Run    &$\overline{m}a^{-1}$ & \multicolumn{2}{c}{$\overline{m}^{\MSbar}$}& &
               $m_sa^{-1}$    & \multicolumn{2}{c}{$m_s^{\MSbar}$}&          &
               $m_ca^{-1}$    & \multicolumn{2}{c}{$m_c^{\MSbar}$}&          \\
        &                  &   NP             &         Pert.            & &
                           &   NP             &         Pert.            & &
                           &   NP             &         Pert.            & \\
\hline
C60a   & 3.45(12)& -       & 4.9(4) & & 78.9(22) & -      & 113(8)&  
          &    -     &  -     &  -  & \\
C60b   & 3.36(9) & -       & 4.8(3) & & 77.3(17) & -      & 110(7)&  
          &    -     &  -     &  -  & \\
C60c   & 3.49(14)& -       & 5.0(4) & & 79.4(25) & -      & 113(8)&  
          &    -     &  -     &  -  & \\
C60d   & 3.21(13)& -       & 4.6(3) & & 74.7(26) & -      & 107(7)&  
          &    -     &  -     &  -   & \\
W60    & 4.36(12) & 5.8(2) & 6.1(8) & & 97.9(20) & 129(4) & 137(17)& 
          & 1335(31) & 1764(62) & 1865(247)  & \\
C62a   & 3.60(29)& -       & 5.2(6) & & 81.1(41) & -      & 117(9)&  
          &  881(18) & -        & 1275(90)  & \\
W62a   & 4.10(18)& 5.6(3)  & 5.9(8) & & 93.3(29) & 127(5) & 134(17)& 
          & 1205(24) & 1646(44) & 1730(212)  & \\
W62b   & 3.98(23)& 5.5(4)  & 5.7(8) & & 91.8(39) & 126(6) & 132(17)& 
          & 1206(32) & 1651(64) & 1738(222)  & \\
W64    & 3.55(18)& 5.1(3)  & 5.3(7) & & 82.0(35) & 119(6) & 123(17)& 
          &    -     &  -  &  -  & \\
C64    & 2.86(17)& -       & 4.3(4) & & 66.9(34) & -       & 100(9)&  
          &    -     &  -  &  - &  \\
\hline
\hline
\end{tabular}
\end{center}
\vspace{0.3truecm}
\caption{Lattice quark masses and the corresponding $\MSbar$ values in 
MeV at NLO from the pseudoscalar meson spectroscopy and 
$a^{-1}$ from $M_{K^*}$. $\MSbar$ masses are at a scale $\mu=2$ 
GeV.\label{tab:quarkm_sp}}
\end{table}

\setlength{\tabcolsep}{.18pc}
\begin{table}[htb]
\begin{center}
\begin{tabular}{||c|cccc|cccc|cccc||}
\hline\hline
 Run    &$\overline{\rho}a^{-1}  $ &\multicolumn{2}{c}{$\overline{m}^{\MSbar}$}& &
         $\rho_s a^{-1} $ &         \multicolumn{2}{c}{$m_s^{\MSbar}$}&          &
         $\rho_c a^{-1}$ &         \multicolumn{2}{c}{$m_c^{\MSbar}$}&        \\
        &                  &   NP             &         Pert.            & &
                           &   NP             &         Pert.            & &
                           &   NP             &         Pert.            & \\
\hline
C60a   & 2.64(11)   & 6.0(3) & 4.0(3) & & 60.4(20) & 136(6) & 93(8) &
       & -        &     -      &  -  & \\
C60b   & 2.52(7)    & 5.7(2) & 3.9(3) & & 58.0(14) & 131(5) & 89(6) &
       & -        &    -       &  -  & \\
C60c   & 2.68(11)   & 6.0(3) & 4.1(3) & & 60.9(20) & 136(6) & 93(7) &
       & -        &    -       &  -  & \\
C60d   & 2.41(12)   & 5.5(3) & 3.7(3) & & 56.1(24) & 128(6) & 86(7) &
       & -        &    -       &  -  & \\
W60    & 3.48(11)   & 5.7(2) & 4.3(5) & & 78.2(20) & 127(5) & 98(11) &
       & 1091(30)  &   1777(67)  & 1362(160) & \\
C62a   & 2.83(23)   & 6.0(5) & 4.4(5) & & 63.6(33) & 134(8) & 99(9) &
       & 820(36)  &   1725(93)  & 1269(113) & \\
W62a   & 3.47(14)   & 5.6(3) & 4.5(5) & & 78.9(23) & 126(5) & 102(11) &
       & 1012(21)  &   1624(43)  & 1311(144) & \\
W62b   & 3.32(19)   & 5.3(3) & 4.3(5) & & 76.5(32) & 124(6) & 99(12) &
       & 1001(29)  &   1610(66)  & 1301(149) & \\
W64    & 3.19(17)   & 4.9(3) & 4.3(5) & & 73.7(33) & 114(6) & 100(13) &
       &  -       &      -     &  -  & \\
C64    & 2.41(14)   & 4.7(3) & 3.8(4) & & 56.3(27) & 110(6) & 90(8) &
       &  -       &      -     &  -  & \\
\hline
\hline
\end{tabular}
\end{center}
\vspace{0.3truecm}
\caption{Lattice quark masses and the 
corresponding $\MSbar$ values at NLO, in MeV
from the Axial Ward Identity and $a^{-1}$ from $M_{K^*}$.
$\MSbar$ masses are at a scale $\mu=2$ GeV.\label{tab:quarkm_wi}}
\end{table}

\section{Conclusions}\label{sec:conclusions}
We have discussed the quark mass renormalization.
We have calculated the quark masses from the meson spectroscopy and from the
Axial Ward Identity using different sets of quenched data 
with $\beta=6.0$, $6.2$ and $6.4$ and using the Wilson and the 
``improved'' SW-Clover action. 
The data at $\beta=6.4$ have been used only for an exploratory 
study as the physical volume and the time extension of the 
lattice may be too small to be reliable.\\
Perturbation theory appears to fail giving inconsistent results for the
two methods. The results for independent NP methods are well
consistent and stable
strongly supporting the reliability of our final results reported in the abstract.
In the $\beta$ range that we have considered and with our
statistics, we do not believe that one can safely extrapolate to the continuum 
limit both for the Wilson and the SW-Clover action.

\section{Appendix}

In this appendix we shall describe the different definition of the lattice strong
coupling constant that we have used for the PT results.

The running coupling constant in $\overline{MS}$ scheme is:
\bea
\frac {\alpha_s(\mu^2)}{4 \pi} =
\frac {1} {\beta_0 ln(\mu^2/\Lambda_{QCD}^2)} \left\{
1 - \frac{\beta_1 ln[ln(\mu^2/\Lambda_{QCD}^2)]}{\beta_0^2
ln(\mu^2/\Lambda_{QCD}^2)} \right\} + \cdots \; ,
\label{eq:alfstr}
\eea

this is related to the bare lattice coupling $\alpha_s^L(a)=g^2_L(a)/4 \pi$
by the relation:
\be \frac{1}{\alpha_s^L(a)}=\frac{1}{\alpha_s(q)}
\left\{ 1+ \frac{\alpha_s(q)}{4\pi}
\Bigl[\beta_0 \ln(\frac{\pi}{qa})^2 + 48.76 \Bigr]\right\}
.\label{eq:alfstrlat} \ee

In ref. \cite{makenzie} a new expansion parameter, $\alpha_s^{V_1}$ has been
introduced based on the heavy quark potential:

\be \frac{1}{\alpha_s^L(a)}=\frac{1}{\alpha_s^{V_1}(q)}
\left\{ 1+\frac{ \alpha_s^{V_1}(q)}{4\pi}
\Bigl[\beta_0 \ln(\frac{\pi}{qa})^2 + 59.09 \Bigr]\right\} ,
\label{eq:alfv}\ee

using the perturbative expansion for the trace of the plaquette one obtains, 
\cite{makenzie}:

\be \frac{1}{\alpha_s^L(a)}=\frac{1}{\alpha_s^{V_1}(\pi/a)
\langle \frac{1}{3}{\rm Tr} U_P\rangle} \Bigl[ 1 +\frac{
\alpha_s^{V_1}(\pi/a)}{4 \pi} 6.45\Bigr] .
\label{eq:plaq2} \ee
 where $U_P$ is the plaquette. The third definition that we have used is:
\be \frac{1}{\alpha_s^L(a)}=\frac{(8 k_c)^4}{\alpha_s^{V_2}} \label{eq:critk}
\ee

where $k_c$ is the critical value of the hopping parameter. 

In our analysis we have used $\alpha_s$, $\alpha_s^{V_1}$, $\alpha_s^{V_2}$ as
they have been used in the literature in the past. One should note that each
definition has different $O(a)$ effects and eq. (\ref{eq:critk}) is only valid at
the LO while eqs. (\ref{eq:alfstr}-\ref{eq:plaq2}) are valid up to NLO. We have
made this choice as it is the same done in ref. \cite{qmass1} and so the
results can immediately be compared.

We have evaluated all cases both in the quenched and unquenched case. In the
latter case the value of $\Lambda_{QCD}^{n_f}$ at the various scales is
obtained from the SLD-LEP result $\Lambda_{QCD}^{5}= 240 \pm 90 ~MeV$
\cite{pdg} by matching the strong coupling constant at $n_f$ and $n_f+1$ at the
threshold of the new flavour \cite{marcia}. In the quenched case we have used 
$\Lambda_{QCD}= 251 \pm 21 ~MeV$ \cite{capit}. Furthermore we have also chosen 
two different scales used to evaluate the strong coupling constant, i.e.
$\alpha_s(\pi/a)$ and $\alpha_s(1/a)$. All these results have been averaged and
the spread taken as the error.
We have also done this analysis using $\Lambda_{QCD} = 340 \pm 120 ~MeV$
and we find that the final average changes by less then $5\%$
which is well within the error. As an example example
$\alpha_s^{\overline{MS}}$ changes from $0.20$ to $0.31$ at a scale
$\mu=1/a=2~GeV$ which leads to a change in the correction of that order.

\begin{ack}
We wish to thank the APE collaboration for allowing us to use the lattice
correlation functions presented here. We warmly thank C.~R.~Allton,
V.~Lubicz, E.~Franco, G.~Martinelli, M.~Testa and A.~Vladikas for
enlightening discussions. We also acknowledge the use of C.~R.~Allton's
analysis program. M.T. acknowledges the support of PPARC through grant
GR/L22744.c
\end{ack}

\newpage
\clearpage

\end{document}